\documentclass[a4paper]{article}
\usepackage[utf8]{inputenc}
\usepackage[UKenglish]{babel}
\usepackage{csquotes}
\usepackage{fullpage}
\usepackage{amsmath, amsthm, amsfonts}
\usepackage{hyperref}
\usepackage{parskip}
\usepackage{xcolor}
\usepackage{graphicx}  
\usepackage{booktabs}  
\usepackage{natbib}
\usepackage{algorithm}
\usepackage[noend]{algorithmic}
\usepackage{subcaption}

\theoremstyle{definition}
\newtheorem{definition}{Definition}[section]

\title{Maximising the Benefits of an Acutely Limited Number of COVID-19 Tests}

\author{
J. Jonnerby\footnote{
Department of Physics, University of Oxford,
\href{mailto:jakob.jonnerby@merton.ox.ac.uk}{jakob.jonnerby@merton.ox.ac.uk}
}
\and 
P. Lazos\footnote{
Department of Computer Science, Sapienza  University  of  Rome \href{mailto:plazos@gmail.com}{lazos@diag.uniroma1.it}
}
\and
E. Lock\footnote{
Department of Computer Science, University of Oxford,
\href{mailto:edwin.lock@merton.ox.ac.uk}{edwin.lock@merton.ox.ac.uk}
}
\and
F. Marmolejo-Coss\'{i}o\footnote{
Department of Computer Science, University of Oxford,
\href{mailto:marmolejo.francisco@gmail.com}{marmolejo.francisco@gmail.com}
}
\and
C. Bronk Ramsey\footnote{
School of Archaeology, University of Oxford,
\href{mailto:christopher.ramsey@arch.ox.ac.uk}{christopher.ramsey@arch.ox.ac.uk}
}
\and
M. Shukla\footnote{
\href{mailto:meghanashukla414@gmail.com}{meghanashukla414@gmail.com}
}
\and
D. Sridhar\footnote{
Department of Zoology, University of Oxford,
\href{mailto:divya.sridhar@merton.ox.ac.uk}{divya.sridhar@merton.ox.ac.uk}
}
}

\date{\today}

\begin{document}

\maketitle

\begin{abstract}
    We propose a novel testing and containment strategy in order to contain the spread of {SARS-CoV2} while permitting large parts of the population to resume social and economic activity. Our approach recognises the fact that testing capacities are severely constrained in many countries. In this setting, we show that finding the best way to utilise this limited number of tests during a pandemic can be formulated concisely as an allocation problem. Our problem formulation takes into account the heterogeneity of the population and uses pooled testing to identify and isolate individuals while prioritising key workers and individuals with a higher risk of spreading the disease. In order to demonstrate the efficacy of our testing and containment mechanism, we perform simulations using a network-based SIR model. Our simulations indicate that applying our mechanism on a population of $100,000$ individuals with only $16$ tests per day reduces the peak number of infected individuals by approximately $20\%$, when compared to the scenario where no intervention is implemented.

\end{abstract}

\section{Introduction}

The role of testing has been intensely discussed during the ongoing COVID-19 pandemic. Testing can be used to identify infected individuals in order to give them appropriate treatment, to trace a chain of transmissions, to estimate the infection rate in the population, and to limit the spread by removing infected individuals from the population through self-isolation. However, the number of tests available is often much less than what would be required to test and isolate all suspected cases. In this resource constrained scenario, tests must to be allocated efficiently in order to have maximum impact on public health. A number of methods to reduce the number of tests required have been discussed, using group testing \citep{Gollier2020} and focusing testing on health care workers \citep{CEPR2020, Grassly2020}. The underlying idea behind both of these approaches is that efficient testing and isolation strategies can be a way to reduce the economic consequences of the pandemic, by allowing individuals who test negative to rejoin the workforce. We propose a more general approach, combining both group testing and segmented testing, as a way of containing the disease while allowing for some economic activity. By considering a reduction in the transmission of the virus to be the principal goal of testing, we formulate an optimisation algorithm that finds the best allocation with a limited testing budget to achieve this goal. 

In group (or pooled) testing, up to $64$ individuals are tested using a single pooled test kit. A pooled test is positive if at least one individual in the pool has the virus, and negative if all individuals in the pool are healthy. This technique has been experimentally verified using PCR methods with SARS-CoV2 \citep{Shental2020, Yelin2020} and synthetic RNA \citep{Ghosh2020}. We propose a simple policy in which everyone who forms part of a group that tests positive is required to self-isolate (or remain in self-isolation), independent of whether they actually have the virus or not. Conversely, anyone forming part of a group that tests negative is permitted to resume normal social and economic activity. This policy is generally applicable and can be used either to ease out of lock-down, or to isolate groups in order to avoid a second peak.

In segmented testing, some segments of the population are given higher testing priority. In our mechanism, we consider segmentation according to an individual's potential exposure to the virus, which can depend on the individual's occupation, age, geographical location, etc. We also take into account a person's cost of isolation; for instance, asking a key worker to self-isolate incurs significant social cost. In our mechanism, prioritising a segment translates into assigning more tests with smaller group sizes to this segment in order to minimise the spread of the virus while minimising the number of individuals self-isolating unnecessarily.

\subsection{Identifying the Exposure of Individuals}
\label{sec:exposure}
Our testing strategy assumes that individuals can be categorised based on their exposure, which is used to determine their risk of contracting and spreading the disease, and their cost of self-isolation. Here we present several approaches that can be taken to estimate the exposure. In practice, the approach used when implementing our testing strategy depends on the information available and the desired level of accuracy when categorising the population.

\begin{itemize}
    \item Social networks: In some cases, it is possible to model society as a network where social relations are represented as connections between individuals. The number of connections of each individual may be used as an indicator of their exposure, and this information could be obtained either from contact tracing, surveys, or from mobile phone proximity tracing applications.
    \item Geographical exposure: An alternative to a social graph approach is to use geolocation data (e.g. from mobile network providers) to identify exposure hot spots, i.e.~areas frequented by many people. Using geolocation data we can determine how many individuals have visited the area and how long they spent there. This creates a heat map of the most heavily visited areas and can assign a higher exposure to people who have visited those areas.
    \item Individual information: In a setting where we have no access to social networks or geographical exposure, we can still make an informed guess about the exposure of an individual based on a combination of simple data points. Possible examples include the age of the individual, the size of the individual's household, whether anyone in the household has shown symptoms, employment status and whether their work puts them in contact with many potentially infected individuals. 
\end{itemize}

\subsection{Optimal test allocation strategy}

To summarise, our proposed strategy takes into account the exposure and cost of self-isolation of different segments of the population in order to determine the group sizes and the number of tests given to each segment, while respecting the limit in the total number of tests available. The tests are then conducted daily, and every individual who forms part of a group that tests positive is required to self-isolate for a set period of time (e.g. for at least $14$ days). This mechanism is general and could be used either during a phased exit from a lock-down period by allowing some people to return to the workforce, provided that they test negative, or to avoid a second peak by monitoring certain segments of the population. 

The following example illustrates the current situation in many countries. Given a population of $100,000$, what is the best way to distribute $10$ tests? Instead of simply testing those that happen to first become ill with severe symptoms, we argue that these tests can be used more efficiently. First of all, group testing is more beneficial than individual testing, as it misses fewer infected individuals. Second of all, certain segments of the population, those with many social interactions such as grocers and nurses, are at high risk of transmitting the virus to others should they become infected and should therefore be prioritised for testing. However, since nurses are needed in health care the social cost of their self-isolation if they are not infected is high, so they should be tested in smaller groups compared to other professions. 

This is a strategic and systematic way to approach the problem, and would greatly reduce the transmission rate compared to randomly testing individuals. Our method works well even for recurring waves of the disease, which has been modelled in the literature \citep{Ferguson2020}, and does not require long-lasting immunity in order to be feasible. It is also possible to include the sensitivity of the test itself as part of the optimisation problem. While we have primarily focused on the ongoing COVID-19 pandemic due to its urgent nature, our approach is generally applicable and can also be used for other infectious diseases.

\section{Formulation of the Optimisation Problem}

\subsection{Notation and Model}
We consider a population denoted by the set $[n] := \{1, \ldots, n \}$. The population is partitioned into $C$ disjoint categories. We denote the $i$-th category by $C_i$ and the number of individuals in $C_i$ by $n_i$ (so $\sum_i n_i = n$). We assume that any individual in category $C_i$ is independently infected with probability $p_i$ (and healthy with probability $q_i = 1- p_i)$. $\mathcal{K}_i$ denotes the set of infected individuals from $C_i$ and $k_i = |\mathcal{K}_i|$ is the number of infected individuals from $C_i$. We also let $\mathcal{K} = \bigcup_i \mathcal{K}_i$ be the set of all infected individuals in the population, and $k = \sum_{i} k_i$ be the number of infected individuals in the population. Finally, each $C_i$ may or may not be in self-isolation. To keep track of this, we associate the parameter $S_i \in \{0,1\}$ as an indicator variable for whether $C_i$ is in self-isolation or not. 

Each individual in $C_i$ has an integral or rational `exposure parameter' $d_i \geq 0$. One possible interpretation of $d_i$ is the number of other people that individual is in regular contact with when not in self-isolation, which corresponds to the number of neighbours they have in a social network. Consequently, a higher value of $d_i$ means that an individual has both a higher probability of being infected in the first place, and in the case of infection, a higher expected number of individuals they can propagate the disease to if they are not in self-isolation. 

All individuals in $C_i$ have a rational cost of self-isolation denoted $\gamma_i \geq 0$. In practical terms, a category consisting of healthcare workers will have a high $\gamma_i$ value, since they are essential in the current crisis, whereas a category with individuals in professions such as software engineering, which are amenable to working from home irrespective of the pandemic, will have a low $\gamma_i$ value. The cost of self-isolation may go beyond how ``essential'' certain professions are. For example, a category consisting of daily wage labourers may have a high $\gamma_i$ value as they do not have the economic means of maintaining self-isolation.

Finally, we introduce a budget constraint on testing kits by assuming that on every given day there are at most $T$ kits available for use in individual or group testing. An optimal solution would most likely utilise all these kits, even though there could be artificial examples where this might not be the case. Moreover, for convenience we define $G=64$, the maximum feasible group size for pooled testing.

\subsection{Lower Granularity Testing}
\label{sec:variable-granularity-testing}

\paragraph{Uniform Group Testing.} Much of the literature in group testing focuses on minimising the number of tests needed to identify all infected individuals in a population (possibly up to a small error) \citep{aldridge19}. In the setting where testing capacities are severely constrained, however, an inherent lack of granularity in pooling tests can actually be an essential feature of the overall testing strategy. In this section, we outline a family of simple group testing protocols along with a corresponding self-isolation policy that balances the objective of both mitigating the spread of the virus and minimising unnecessary self-isolation.

\begin{definition}[Uniform group testing]
For a given population of $n$ individuals, we say a testing protocol is a uniform group test of granularity $g \geq 1$ and scope $\ell \geq 1 $ if it performs group tests on $\ell$ {\em disjoint} groups of size $g$. Notice that this testing strategy uses $\ell$ tests and that it must necessarily be the case that $\ell g \leq n$. 
\end{definition}

As mentioned in the previous section, our population is segmented into $C$ different categories according to the exposure, self-isolation costs of individuals, and whether $C_i$ is self-isolating or not. With this segmentation in hand, our testing strategy is straightforward: for each category $C_i$, we perform uniform group testing of granularity $g_i$ and scope $\ell_i$. Since the cost of performing each uniform group test is $\ell_i$, the overall testing protocol for the whole population has to respect $\sum_{i=1}^C \ell_i \leq T$, as this is the testing budget of the system. 

Now that we have outlined a family of feasible testing protocols, we must specify both a containment protocol that is implemented upon gathering results, as well as an overall social objective that the combination of the testing and containment is aiming to achieve. This will allow us to select the optimal testing protocol for each population category.

\paragraph{Containment Protocol.} Our containment protocol can be described in terms of what it recommends for self-isolated and non-self-isolated segments of the population.
\begin{itemize}
    \item Suppose that $C_i$ is a segment of the population not under self-isolation (i.e. $S_i = 0)$. If an individual belongs to a group that tests positive, they self-isolate, otherwise they are not obliged to do so.
    \item Suppose that $C_i$ is a segment of the population under self-isolation (i.e. $S_i = 1)$. If an individual belongs to a group that tests negative, they are released from self-isolation.
\end{itemize}

\paragraph{Objective.} Given this setting, our objective is twofold:
\begin{itemize}
    \item Amongst individuals who are not in self-isolation, minimising the number of infected people that are not tested, in order to suppress transmission, with a greater emphasis on containing highly connected individuals (i.e. which belong to a highly connected category). 
    \item Mitigating the impact of healthy people
    in positive groups who must subsequently self-isolate unnecessarily.
\end{itemize}

\subsection{Formalising the Optimisation Problem}
\label{sec:optimisation-details}
We first formalise the objective that was stated informally above. Recall that $n_i$ denotes the number of agents in category $C_i$. Suppose that we perform uniform testing of granularity $g_i$ and scope $\ell_i$ on each category $C_i$. It follows that we apply $\ell_i$ tests to each population category. Similarly, we also define $r_i = n_i - g_i \ell_i$ as the number of individuals in category $i$ that are untested.

\subsubsection{Segments not under Self-Isolation}

We begin by considering a category $C_i$ that is not under self-isolation. 

\paragraph{Untested Individuals}
We recall that our testing strategy may have $r_i$ untested individuals within category, $C_i$. If these individuals are healthy, they do not incur a cost of self-isolation, as we have assumed that $S_i = 0$. On the other hand, if any such individual is infected, they may infect new individuals according to their exposure.  By assumption, each untested individual is infected with probability $p_i$. Hence, the expected number of infected individuals in $C_i$ that are not detected is $p_i r_i$. As the number of people an individual infects is proportional to $d_i$, we multiply the expected number of infected individuals by $d_i$ to obtain the cost incurred from untested individuals:
$$
    d_i p_i r_i.
$$

\paragraph{Healthy Individuals in Positive Tests}
Our second objective is to minimise the number of healthy individuals who are incorrectly told to self-isolate. Suppose we perform uniform group testing of granularity $g_i$ and $\ell_i$ on category $C_i$. As each individual is healthy with probability $q_i = 1-p_i$, the probability that a test is positive is given by $1-q_i^{g_i}$,
while the expected number of healthy individuals in a group of size $g_i$ conditioned on the test being positive is $g_i -  {g_i p_i}{(1-q_i^{g_i})}$. Hence by multiplying these two terms, we get the expected number of healthy individuals that are self-isolating unnecessarily,
\begin{equation*}
    g_i(q_i - q_i^{g_i}).
\end{equation*}

Each unnecessary self-isolation of a member of category $C_i$ incurs a social cost of $\gamma_i$ by assumption, and since there are $\ell_i$ groups tested overall, the total cost of unnecessary self-isolation for the category is 
$$
\gamma_i \ell_i g_i(q_i - q_i^{g_i}).
$$

\paragraph{Overall Segment Loss}

We put the above expressions together to formulate a preliminary loss, $L_i'$, incurred by $C_i$ under the choice, $g_i,\ell_i$ of granularity and scope of testing within $C_i$. 
$$
L_i'(g_i,\ell_i,n_i) = d_i p_i r_i + \gamma_i \ell_i g_i(q_i - q_i^{g_i})
$$
If we make the substitution $r_i = n_i - g_i 
\ell_i$, the above expression becomes:
$$
L_i'(g_i,\ell_i,n_i) = d_i p_i n_i + \ell_i((\gamma_i q_i - d_i p_i)g_i - \gamma_i g_i q_i^{g_i}).
$$
The left-most term is a constant that does not depend on the choice of $g_i$ and $\ell_i$, hence we can ignore it in loss minimisation. Furthermore, we also define
$$
\theta_i(x) = (\gamma_i q_i - d_i p_i)g_i - \gamma_i g_i q_i^{g_i}
$$
so that we obtain an equivalent loss function:
$$
L_i(g_i,\ell_i) = \ell_i \theta_i(g_i)
$$

\subsubsection{Segments under Self-Isolation} 

We now consider a category $C_i$ that is under self-isolation.

\paragraph{Untested Individuals}
Once more, we focus on the $r_i$ untested individuals for a given testing regime. Since $C_i$ is currently under self-isolation, if an individual is infected they cannot spread the infection further, so there is no cost incurred for new cases attributed to that individual. On the other hand, if an untested individual is healthy, we must incur the cost of prolonging their unnecessary self-isolation. By assumption, each untested individual is healthy with probability $q_i$, and if untested, this individual remains under self-isolation and incurs a cost of $
\gamma_i$. Consequently, the cost of untested individuals is

$$
\gamma_i q_i r_i.
$$

\paragraph{Healthy Individuals in Positive Tests}
Once again, our second objective is to minimise the number of healthy individuals who are incorrectly told to self-isolate. Via an identical analysis to the case where $C_i$ is not under self-isolation, we obtain the following total cost from healthy individuals unnecessarily told to keep self-isolating by being involved in a positive group test. 

$$
\gamma_i \ell_i g_i(q_i - q_i^{g_i}).
$$

\paragraph{Overall Segment Loss}

We put the above expressions together to formulate a preliminary loss, $L_i'$, incurred by $C_i$ under the choice $g_i,\ell_i$ of granularity and scope of testing within $C_i$: 
$$
L_i'(g_i,\ell_i,n_i) = \gamma_i q_i r_i + \gamma_i \ell_i g_i(q_i - q_i^{g_i}).
$$
If we make the substitution $r_i = n_i - g_i 
\ell_i$, the above expression becomes:
$$
L_i'(g_i,\ell_i,n_i) = \gamma_i q_i n_i + \ell_i( - \gamma_i g_i q_i^{g_i}).
$$
The left-most term is a constant that does not depend on the choice of $g_i$ and $\ell_i$, hence we can ignore it in loss minimisation. Furthermore, we also define
$$
\theta_i(x) = - \gamma_i g_i q_i^{g_i},
$$
so that we obtain an equivalent loss function:
$$
L_i(g_i,\ell_i) = \ell_i \theta_i(g_i)
$$

\subsubsection{The Overall Optimisation Problem}

In the previous section we've done all the groundwork to see that for a given choice of $g_i$ and $\ell_i$ per each $C_i$, the overall loss of the testing allocation and containment strategy can be expressed as:
$$
L(g,\ell) = \sum_{i=1}^C \ell_i \theta_i (g_i)
$$
Clearly this objective function is not only separable (as the sum of individual loss functions per segment), but linear in $\ell_i$, provided that $g_i$ is constant. With this in hand, we can formulate our optimisation program as an integer linear programme (ILP) in $\ell$ with $C$ free variables and $C+1$ constraints.

\begin{equation}
\label{eq:opt2}
    \begin{aligned}
        &\underset{g,\ell}{\text{min}}&&
        \sum_{i=1}^C \ell_i \theta_i(g_i)  \\
        &\text{subject to} &&\ell_i \leq  \frac{n_i}{g_i} \\
        & &&\sum_{i=1}^C \ell_i \leq T \\
        & &&g_i \leq G \\
        & && g_i, \ell_i \in \mathbb{N} \\        
    \end{aligned}
\end{equation}

This suggests a simple approach: for every feasible group size vector $g$, solve $\eqref{eq:opt2}$ and return the combination of $g$ and $l$ that minimises the objective. More importantly, the separable nature of $\eqref{eq:opt2}$ implies that the optimal allocation of the $T$ tests can be computed via a simple greedy algorithm, which we outline in Algorithm~\ref{alg:OSUGT}. As there are at most $G^C$ different ways to fix $g$, this method works in practice if we have few population categories and limit the different group sizes we use to test each category.

\begin{algorithm} 
\caption{Optimal Segmented Uniform Group Testing}
\label{alg:OSUGT}                           
\begin{algorithmic}[1]                    
    \REQUIRE 
    \STATE Granularity Range: $R_G \subseteq [G]^C$ 
      \item[\textbf{Iterating over Granularities:}]
    \STATE $OPT \leftarrow \infty$
    \FOR{$g \in R_G$}
        \STATE $\ell_i \leftarrow 0$ for $i \in [C]$
        \STATE Compute $\sigma$, an ordering of $C_i$ with respect to increasing $\theta_i(g_i)$ values
        \STATE $T_r \leftarrow T$
        \STATE $i \leftarrow 1$
        \WHILE{$T_r > 0$}
            \STATE $\ell_{\sigma(i)} \leftarrow \min \{T_r, \Big \lfloor \frac{n_{\sigma(i)}}{g_{\sigma(i)}} \Big \rfloor \}$
            \STATE $T_r \leftarrow T_r - \ell_{\sigma(i)}$
            \STATE $i \leftarrow i+1$
        \ENDWHILE
        \IF{$\sum_{i=1}^C \ell_i \theta_i(g_i) < OPT$}
            \STATE $OPT \leftarrow \sum_{i=1}^C \ell_i \theta_i(g_i)$
            \STATE $g^*,\ell^* \leftarrow g,\ell$
        \ENDIF
                
    \ENDFOR   
    \STATE
    \RETURN $g^*,\ell^*,OPT$
  
\end{algorithmic}
\end{algorithm}

\subsection{Alternative Optimisation Algorithms}

Algorithm $\ref{alg:OSUGT}$ fixes granularities and computes optimal testing allocations, hence it is only efficient if the population has few segmentations and group sizes. An alternative approach to solving $\eqref{eq:opt2}$ lies in trying precisely the opposite approach, whereby for fixed $\ell_i$, we compute the optimal value of $\theta_i(g_i)$ under the constraints that $1 \leq g_i \leq  \min \{G, \frac{n_i}{\ell_i}  \}$. Let $D_i = [1,  \frac{n_i}{\ell_i} ] \cap \mathbb{Z}$, then we can define 
$$
\bar{x}(\ell_i) = \text{argmin}_{x \in D_i} \theta_i(x)
$$
Consequently, the optimisation can rid itself of the dependence on $g_i$, and instead minimise $\sum_{i=1}^C f_i(\ell_i)$, where $f_i(\ell_i) = \ell_i \theta_i(\bar{x}(\ell_i))$. This of course is still subject to the fact that testing allocations must respect the given budget $T$, i.e. $\sum_{i=1}^C \ell_i \leq T$. In other words, we get the following optimisation problem: 

\begin{equation}
\label{eq:opt3}
    \begin{aligned}
        &\underset{\ell}{\text{min}}&&
        \sum_{i=1}^C f_i(\ell_i)  \\
        &\text{subject to} && \sum_{i=1}^C \ell_i \leq T  \\
        & && \ell_i \in \mathbb{N} \\        
    \end{aligned}
\end{equation}

The structure of $f_i$ as a function of $\ell_i$ is not directly amenable to optimisation. That being said, it can be described efficiently, which opens the possibility to black-box optimisation, or simply an exhaustive search over all possible testing allocations, namely test vectors that satisfy $\sum_{i=1}^C \ell_i \leq T$. 

This latter option can be very useful in the regime where testing allocation can only be done in terms of bundles. In particular, suppose that the budget of $T$ tests consists of $B$ bundles of $T/B$ tests. Then, the number of possible allocations of the $B$ bundles to $C$ categories is $\binom{B+1}{C-1}$, which can be more efficient in the regime where $B < G$. We describe this allocation procedure in Algorithm \ref{alg:bundle-allocation}. In particular, we let $R_B$ be the set of all possible testing allocations $\ell  = (\ell_i)_{i=1}^C$ that are composed of bundles of $T/B$ tests in a given category $C_i$, such that the overall testing allocation remains feasible.

\begin{algorithm} 
\caption{Bundle Allocation Enumeration}
\label{alg:bundle-allocation}                      
\begin{algorithmic}[1]                    
    \REQUIRE 
    \STATE Bundled Testing Range: $R_B$

      \item[\textbf{Iterating over Allocations:}]
    \STATE $OPT \leftarrow \infty$
    \FOR{$\ell \in R_B$}
        \IF{$\sum_{i=1}^C f_i (\ell_i) < OPT$}
            \STATE $OPT \leftarrow \sum_{i=1}^C f_i (\ell_i)$
            \STATE $\ell^* \leftarrow \ell$
        \ENDIF
                
    \ENDFOR   
    \RETURN $\ell^*,OPT$
  
\end{algorithmic}
\end{algorithm}

\subsection{Adjusting Priorities in the Objective}
Here we describe a simple adjustment to the optimisation object that gives policy makers a way to balance the two objectives of reducing the spread of the virus and maintaining economic activity. In previous sections we noticed that the loss of a given segment was different depending on whether the segment was in self-isolation or not. In the case of $C_i$ being under self-isolation, we obtained a preliminary segment loss of 
$$
L_i'(g_i,\ell_i,n_i) = d_i p_i r_i + \gamma_i \ell_i g_i(q_i - q_i^{g_i})
$$
and in the case of $C_i$ not being under self-isolation, we obtained a preliminary segment loss of 
$$
L_i'(g_i,\ell_i,n_i) = \gamma_i q_i r_i + \gamma_i \ell_i g_i(q_i - q_i^{g_i})
$$

We note that a policy-maker may have different priorities in terms of whether they wish to mitigate virus spread or reduce the impact of unnecessary self-isolation. We briefly demonstrate that our model is general enough to encompass different priorities. This can be done by introducing a balancing parameter $\beta \in [0,1]$. Larger values of $\beta$ prioritise the suppression of the virus, while smaller values allow greater numbers of individuals to resume economic activity at the expensive of containing the virus to a lesser extent.

In order to incorporate the balancing parameter $\beta$, we reformulate our preliminary losses for when $C_i$ is under self-isolation as
$$
L_i'(g_i,\ell_i,n_i) = \beta (d_i p_i r_i) + (1-\beta)\gamma_i \ell_i g_i(q_i - q_i^{g_i})
$$
and in the case of $C_i$ not being under self-isolation, 
$$
L_i'(g_i,\ell_i,n_i) = (1-\beta) \left( \gamma_i q_i r_i + \gamma_i \ell_i g_i(q_i - q_i^{g_i}) \right)
$$
Ultimately, the same goal could be achieved by simply letting $d_i' = \beta d_i$ and $\gamma_i' = (1-\beta)\gamma_i$ in our original optimisation program.

\section{Modelling Testing Allocation During an Epidemic Process}

\begin{figure}[h]
\centering
\begin{subfigure}[t]{.5\textwidth}
  \centering
  \includegraphics[width=0.95\textwidth]{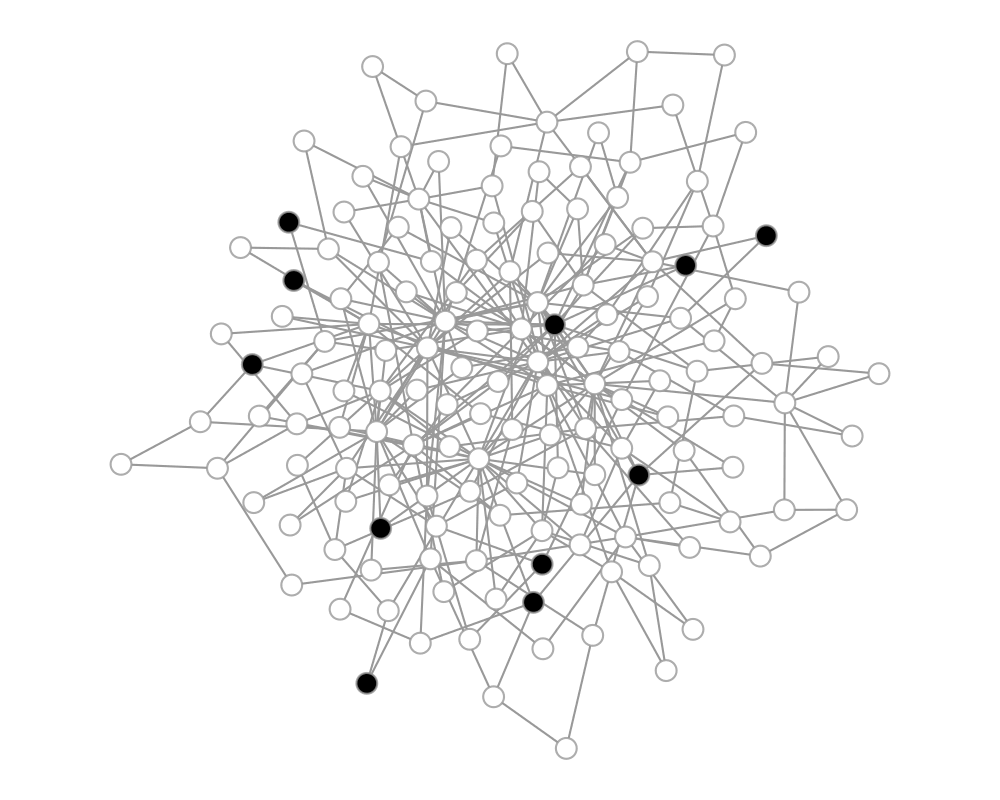}
  \caption{}
  \label{fig:network}
\end{subfigure}%
\begin{subfigure}[t]{.5\textwidth}
  \centering
  \includegraphics[width=.95\textwidth]{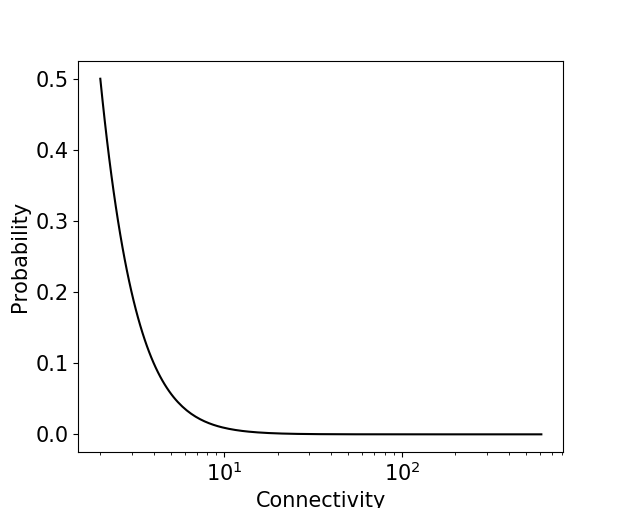}
  \caption{}
  \label{fig:connectivity}
\end{subfigure}
\caption{(a) Example of a scale-free network of $150$ nodes generated using the Barabási–Albert method. The white nodes are susceptible and the black nodes are initially infected, chosen at random each simulation. (b) Probability density function of the connectivity of a node in a network of $100,000$ nodes.}
\label{fig:test}
\end{figure}

In order to verify our testing allocation mechanism, we developed a simple network-based Susceptible-Infected-Recovered (SIR) model on a heterogeneous population using the graph-tool Python library \citep{graph_tool}. We used this to model the impact of an uneven distribution of the exposure on the epidemiological process. The exposure parameter $d_i$ was identified with the connectivity (also known as the degree) of a node in the network model - i.e.~the number of connections to other nodes (neighbours) in the network. The Barabási–Albert model was used to generate a scale-free network with each node having at least two connections, (see Figure~\ref{fig:network}) and the number of connections distributed according to a power law $P(k) \sim k^{-3}$ (see Figure \ref{fig:connectivity}) \citep{RevModPhys.74.47}.

The SIR model is run for a fixed number of (discrete) time steps. Assuming that each time corresponds to one day, our simulation runs for 200 steps corresponding to 200 days. At each time step, an infected node recovers with probability $\gamma$.
If the infected node is not self-isolating, it also infects each susceptible neighbouring node with probability $\beta$. The intervention mechanism consists of performing a fixed number of tests each day, either on groups or individuals, and enforcing social isolation for everyone in a group that tests positive for the virus. The parameters $\beta=0.02$ and $\gamma=0.0427$ were chosen such that average number of secondary infections and time until recovery are $R_0\sim2.5$ \citep{Flaxman2020} and $t_\text{recovery}=14$ days \citep{Lauer2020}. Furthermore, $20 \%$ of the individuals in the model were assumed to be key workers, for whom self-isolation would be very costly. The probability of being a key worker was weighted using the logarithm of the connectivity, in order to account for the fact that some key workers are more likely to have a large number of connections (e.g. nurses, grocers): 
$$P_\text{key worker} = r\frac{\log(K)}{\sum\log(K)},$$
where $r$ is the fraction of key workers in the population. We explored the impact of different testing allocation scenarios on the epidemic outcome, paying particular attention to the peak number of infected individuals and to the number of quarantined individuals at any given point in time. The scenarios were constructed with the testing capacity of the United Kingdom in mind, approximately $10,000$/day ($16$/day per $100,000$ inhabitants). Testing was implemented using two different strategies. In the `optimised' testing strategy, the population was divided in three segments depending on their connectivity. All tests were focused on the highest connectivity segment (those with a daily number of connections greater than $6$). Half the tests were distributed to key workers within the segment, who were tested individually, and the other half were used to test individuals in groups of $10$. In the `random sampling' testing strategy, testing was conducted in groups of $10$, independently of an individual's connectivity or key worker status.

We assumed an initial infection rate of $0.1 \%$ in the population and that testing started from day $10$. Since the temporal resolution of our model was already limited to one day, we automatically include effects of delayed testing outcomes. For each scenario, a population of $100,000$ was simulated over $200$ days, and was initiated at random $100$ times, which allowed us to obtain the mean and the standard deviation of our simulations.

\subsection{Results}

\begin{figure}[h]
    \centering
    \includegraphics[width=0.7\textwidth]{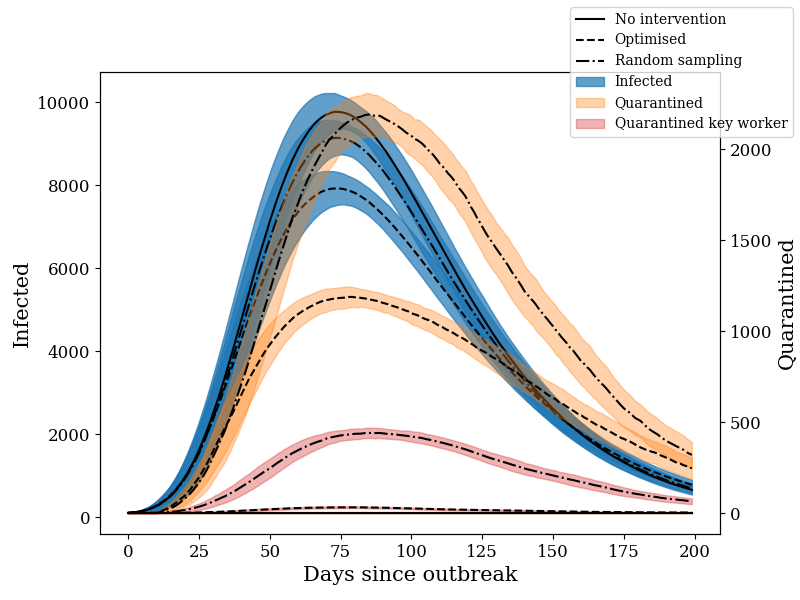}
    \caption{Comparison of the impact on the number of infected (left axis), number of quarantined individuals and number of quarantined key workers (right axis) during an epidemic process, assuming no other interventions other than testing and self-isolation of confirmed positive cases. A population of $100,000$ individuals was simulated and $16$ tests were performed each day, either testing groups of $10$ at random or by segmenting the population and testing only individuals with more than $6$ connections.}
    \label{fig:SIR_simulation}
\end{figure}  

In our model, the total number of infected individuals at any one point was found to be at maximum $10,000$, or $10 \%$ of the population. The `optimised' testing and isolation strategy shows a reduction of $(19\pm5.5) \%$  of the peak height in the number of infected compared to the baseline scenario. The random sampling strategy reduced the peak height by $(6.5 \pm 6) \%$ compared to the baseline. The simulation also shows that fewer people need to be quarantined using the optimised scenario compared to random sampling: a $(45 \pm 3.8) \%$ reduction in the number of people self-isolating during the peak. The peak number of quarantined key workers is reduced by $(93 \pm 1.2) \%$ compared to that of the `random sampling' strategy.

\section{Discussion}

Testing is a valuable but limited resource that can have many different purposes during an pandemic. In this study, we have considered methods of maximising the utility of a limited number of tests in a heterogeneous population. By combining segmented testing with group testing, our proposed testing mechanism can dramatically enhance the utility of a small number of test kits.
Our approach is conceptually simple and general. It can be applied in scenarios where countries want to ease out of lock-down, as well as to avoid a possible second peak in the number of infections.

In order to validate our method, we have conducted simulations on a network-based SIR model. The results show that our strategy can substantially reduce the peak height in the number of infections even when the number of tests available is severely constrained. This is especially valuable in large outbreaks, where testing capacities are significantly smaller than the affected population. Our simulations do not include implementing any other interventions, such as isolating symptomatic cases, which would further increase the reach of a limited number of tests by mainly testing asymptomatic individuals.

Importantly, our method can be used together with a contact tracing app. Assuming that information could be obtained about each individual's connectivity from the app, testing resources could be focused on those with high connectivity.

\subsection{Future Directions}
Our optimisation problem from Section \ref{sec:optimisation-details} has already proved useful in providing non-trivial testing strategies over a segmented population, but there are multiple natural refinements to the model.

\paragraph{Imperfect Testing}
A key assumption in our testing regime is that tests are accurate, but in practice, either due to inherent limitations of testing technology, or even human error, this is not the case. In reality, tests have a false positive and false negative rate which need to be taken into account when deciding whether individuals self-isolate or not as a result of test results. To model this scenario within our resource-constrained optimisation framework, we can assume that a given group of $g$ individuals is given a test of ``intensity'' $s$, and as a function of the intensity, the test has a false positive rate of $p^+(s)$ and a false negative rate of $p^-(s)$. Intensity can for example represent making $s$ independent tests on the same group, and taking the majority vote of the results as the final group test result. In this setting, we can consider an identical containment policy to what we've outlined in Section \ref{sec:optimisation-details}, but refine the relevant costs with more nuanced false positive and false negative rates. With this in hand, we obtain a similar optimisation program where each segment of the population, $C_i$ is tested at granularity $g_i$, scope $\ell_i$ and intensity $s_i$, resulting in a testing cost of $\ell_i s_i$ for that segment. 

\paragraph{Pseudo-Group Testing}
In practice, performing group testing as prescribed might still be prohibitively expensive or complicated; even though the underlying test kit usage remains unchanged between typical testing regimes and systematic group testing regimes, as group sizes increase, logistics scale with the reach of the testing regime. In particular, swabs still need to be collected in the same fashion and in addition, group tests become more unreliable as group sizes increase. 

A possible alternative is to \emph{simulate} group testing using randomisation: within a group, only a random subset (that could contain just one individual) is tested. This approach cannot be immediately used if the groups themselves are randomly selected. Especially if the probability that an individual is infected is small, for a group to test positive there would need to be an infected individual inside it that is \emph{also} randomly selected, drastically changing the result from proper group testing. This may or may not be desirable, as although different, random group testing can (on a large scale) reflect more accurately the situation within groups: the probability that a random group (of any size) testing positive is exactly $p_i$. However, pseudo-group testing might be appropriate for groups which are highly correlated and a randomly selected individual is most likely a good representative of the group.

\paragraph{Non-Disjoint Group Tests}
In the mechanism above, we perform uniform group testing of scope $\ell_i$ and granularity $g_i$ to segment $C_i$, which corresponds to testing $\ell_i$ \textit{disjoint} groups of size $g_i$. In practice, ensuring the disjointness of groups that are tested can be challenging. Instead, we can consider the simpler approach of randomly sampling groups of size $g_i$ with replacement.

It is important to note that this approach may result in individuals being tested more than once in a single round, leading to `testing fatigue'. Moreover, given the same number of tests, this approach yields weakly less information than uniform group testing. Indeed, the expected utility derived from tests scales sub-linearly with the number of tests. However, non-disjoint group testing may be a conceptually simpler approach when dealing with the increased number of tests required in the setting with imperfect testing, as it allows us to increase the number of tests per segment beyond $n_i / g_i$.

\section{Acknowledgements}
We would like to thank Sandeep Krishna and Manoj Gopalkrishnan for their valuable input. We also acknowledge the helpful advice and comments from members of the Mechanism Design for Social Good (MD4SG) initiative. Finally, we thank Alan Garfinkel and Paul Klemperer for their feedback.

\bibliography{library}

\begin{thebibliography}{12}
\providecommand{\natexlab}[1]{#1}
\providecommand{\url}[1]{\texttt{#1}}
\expandafter\ifx\csname urlstyle\endcsname\relax
  \providecommand{\doi}[1]{doi: #1}\else
  \providecommand{\doi}{doi: \begingroup \urlstyle{rm}\Url}\fi

\bibitem[Albert and Barab\'asi(2002)]{RevModPhys.74.47}
R.~Albert and A.-L. Barab\'asi.
\newblock Statistical mechanics of complex networks.
\newblock \emph{Rev. Mod. Phys.}, 74:\penalty0 47--97, Jan 2002.
\newblock \doi{10.1103/RevModPhys.74.47}.
\newblock URL \url{https://link.aps.org/doi/10.1103/RevModPhys.74.47}.

\bibitem[Aldridge et~al.(2019)Aldridge, Johnson, and Scarlett]{aldridge19}
M.~Aldridge, O.~Johnson, and J.~Scarlett.
\newblock {Group Testing: An Information Theory Perspective}.
\newblock \emph{Foundations and Trends{\textregistered} in Communications and
  Information Theory}, 15\penalty0 (3-4):\penalty0 196--392, 2019.
\newblock ISSN 1567-2190.
\newblock \doi{10.1561/0100000099}.
\newblock URL \url{http://dx.doi.org/10.1561/0100000099}.

\bibitem[Ferguson et~al.(2020)Ferguson, Laydon, Nedjati-Gilani, Imai, Ainslie,
  Baguelin, Bhatia, Boonyasiri, Cucunub{\'{a}}, Cuomo-Dannenburg, Dighe,
  Dorigatti, Fu, Gaythorpe, Green, Hamlet, Hinsley, Okell, {Van Elsland},
  Thompson, Verity, Volz, Wang, Wang, {Gt Walker}, Walters, Winskill,
  Whittaker, Donnelly, Riley, and Ghani]{Ferguson2020}
N.~M. Ferguson, D.~Laydon, G.~Nedjati-Gilani, N.~Imai, K.~Ainslie, M.~Baguelin,
  S.~Bhatia, A.~Boonyasiri, Z.~Cucunub{\'{a}}, G.~Cuomo-Dannenburg, A.~Dighe,
  I.~Dorigatti, H.~Fu, K.~Gaythorpe, W.~Green, A.~Hamlet, W.~Hinsley, L.~C.
  Okell, S.~{Van Elsland}, H.~Thompson, R.~Verity, E.~Volz, H.~Wang, Y.~Wang,
  P.~{Gt Walker}, C.~Walters, P.~Winskill, C.~Whittaker, C.~A. Donnelly,
  S.~Riley, and A.~C. Ghani.
\newblock {Report 9: Impact of non-pharmaceutical interventions (NPIs) to
  reduce COVID-19 mortality and healthcare demand}.
\newblock \emph{Imperial College London}, 2020.
\newblock \doi{10.25561/77482}.
\newblock URL \url{https://doi.org/10.25561/77482}.

\bibitem[Flaxman et~al.(2020)Flaxman, Mishra, Gandy, Unwin, Coupland, Mellan,
  Zhu, Berah, Eaton, Guzman, Schmit, Callizo, Ainslie, Baguelin, Blake,
  Boonyasiri, Boyd, Cattarino, Ciavarella, Cooper, Cucunubá, Cuomo-Dannenburg,
  Dighe, Djaafara, Dorigatti, van Elsland, FitzJohn, Fu, Gaythorpe, Geidelberg,
  Grassly, Green, Hallett, Hamlet, Hinsley, Jeffrey, Jorgensen, Knock, Laydon,
  Nedjati-Gilani, Nouvellet, Parag, Siveroni, Thompson, Verity, Volz, Walker,
  Walters, Wang, Wang, Watson, Whittaker, Winskill, Xi, Ghani, Donnelly, Riley,
  Okell, Vollmer, Ferguson, and Bhatt]{Flaxman2020}
S.~Flaxman, S.~Mishra, A.~Gandy, H.~J.~T. Unwin, H.~Coupland, T.~A. Mellan,
  H.~Zhu, T.~Berah, J.~W. Eaton, P.~N.~P. Guzman, N.~Schmit, L.~Callizo,
  K.~E.~C. Ainslie, M.~Baguelin, I.~Blake, A.~Boonyasiri, O.~Boyd,
  L.~Cattarino, C.~Ciavarella, L.~Cooper, Z.~Cucunubá, G.~Cuomo-Dannenburg,
  A.~Dighe, B.~Djaafara, I.~Dorigatti, S.~van Elsland, R.~FitzJohn, H.~Fu,
  K.~Gaythorpe, L.~Geidelberg, N.~Grassly, W.~Green, T.~Hallett, A.~Hamlet,
  W.~Hinsley, B.~Jeffrey, D.~Jorgensen, E.~Knock, D.~Laydon, G.~Nedjati-Gilani,
  P.~Nouvellet, K.~Parag, I.~Siveroni, H.~Thompson, R.~Verity, E.~Volz, P.~G.
  Walker, C.~Walters, H.~Wang, Y.~Wang, O.~Watson, C.~Whittaker, P.~Winskill,
  X.~Xi, A.~Ghani, C.~A. Donnelly, S.~Riley, L.~C. Okell, M.~A.~C. Vollmer,
  N.~M. Ferguson, and S.~Bhatt.
\newblock {Estimating the number of infections and the impact of
  non-pharmaceutical interventions on COVID-19 in 11 European countries}.
\newblock \emph{Imperial College London}, March:\penalty0 1--35, 2020.
\newblock \doi{10.25561/77731}.
\newblock URL \url{https://doi.org/10.25561/77731}.

\bibitem[Ghosh et~al.(2020)Ghosh, Rajwade, Krishna, Gopalkrishnan, and
  Schaus]{Ghosh2020}
S.~Ghosh, A.~Rajwade, S.~Krishna, N.~Gopalkrishnan, and T.~E. Schaus.
\newblock {Tapestry : A Single-Round Smart Pooling Technique for COVID-19
  Testing}.
\newblock \emph{medRxiv}, 2020.
\newblock URL \url{Ghosh2020}.

\bibitem[Gollier and Gossner(2020)]{Gollier2020}
C.~Gollier and O.~Gossner.
\newblock {Group Testing Against Covid-19}.
\newblock \emph{EconPol Policy Brief}, 2020.

\bibitem[Grassly et~al.(2020)Grassly, Pons-salort, Parker, White, Ainslie, and
  Baguelin]{Grassly2020}
N.~C. Grassly, M.~Pons-salort, E.~P.~K. Parker, P.~J. White, K.~Ainslie, and
  M.~Baguelin.
\newblock {Report 16 : Role of testing in COVID-19 control}.
\newblock \emph{Imperial College London}, April:\penalty0 1--13, 2020.

\bibitem[Lauer et~al.(2020)Lauer, Grantz, Bi, Jones, Zheng, Meredith, Azman,
  Reich, and Lessler]{Lauer2020}
S.~A. Lauer, K.~H. Grantz, Q.~Bi, F.~K. Jones, Q.~Zheng, H.~R. Meredith, A.~S.
  Azman, N.~G. Reich, and J.~Lessler.
\newblock {The Incubation Period of Coronavirus Disease 2019 (COVID-19) From
  Publicly Reported Confirmed Cases: Estimation and Application}.
\newblock \emph{Annals of Internal Medicine}, 03 2020.
\newblock ISSN 0003-4819.
\newblock \doi{10.7326/M20-0504}.
\newblock URL \url{https://doi.org/10.7326/M20-0504}.

\bibitem[{Matthew Cleevely} et~al.(2020){Matthew Cleevely}, Susskind, Vines,
  Vines, and Wills]{CEPR2020}
{Matthew Cleevely}, D.~Susskind, D.~Vines, L.~Vines, and S.~Wills.
\newblock {A workable strategy for Covid-19 testing: Stratified periodic
  testing rather than universal random testing}.
\newblock \emph{CEPR Press}, 8, 2020.

\bibitem[{Peixot, Tiago}(2020)]{graph_tool}
{Peixot, Tiago}.
\newblock Graph tool, 2020.
\newblock URL \url{https://graph-tool.skewed.de/}.
\newblock [Online; accessed 15-April-2020].

\bibitem[Shental et~al.(2020)Shental, Levy, Skorniakov, Wuvshet, Shemer-Avni,
  Porgador, and Hertz]{Shental2020}
N.~Shental, S.~Levy, S.~Skorniakov, V.~Wuvshet, Y.~Shemer-Avni, A.~Porgador,
  and T.~Hertz.
\newblock {Efficient high throughput SARS-CoV-2 testing to detect asymptomatic
  carriers}.
\newblock \emph{medRxiv}, 2020.
\newblock \doi{10.1101/2020.04.14.20064618}.
\newblock URL
  \url{https://www.medrxiv.org/content/early/2020/04/20/2020.04.14.20064618}.

\bibitem[Yelin et~al.(2020)Yelin, Aharony, Shaer-Tamar, Argoetti, Messer,
  Berenbaum, Shafran, Kuzli, Gandali, Hashimshony, Mandel-Gutfreund,
  Halberthal, Geffen, Szwarcwort-Cohen, and Kishony]{Yelin2020}
I.~Yelin, N.~Aharony, E.~Shaer-Tamar, A.~Argoetti, E.~Messer, D.~Berenbaum,
  E.~Shafran, A.~Kuzli, N.~Gandali, T.~Hashimshony, Y.~Mandel-Gutfreund,
  M.~Halberthal, Y.~Geffen, M.~Szwarcwort-Cohen, and R.~Kishony.
\newblock {Evaluation of COVID-19 RT-qPCR test in multi-sample pools}.
\newblock \emph{medRxiv}, 2020.
\newblock \doi{10.1101/2020.03.26.20039438}.
\newblock URL
  \url{https://www.medrxiv.org/content/early/2020/03/27/2020.03.26.20039438}.

\end{thebibliography}

\appendix

\end{document}